\documentclass[fleqn,usenatbib]{mnras}

\usepackage{newtxtext,newtxmath}
\usepackage{hyperref}
\usepackage[T1]{fontenc}
\usepackage{ae,aecompl}
\usepackage{subfig}
\usepackage{enumerate}
\usepackage{graphicx}	
\usepackage{amsmath}	
\usepackage{amssymb}	
\usepackage{float}
\usepackage{xspace} 
\usepackage[nameinlink,capitalise]{cleveref}
\usepackage[dvipsnames]{xcolor}

\newcommand{\ziangtxt}[1]{{\color{black}{{#1}}}}

\title[Galaxy cluster mass CNN]{Galaxy cluster mass estimation with deep learning and hydrodynamical simulations}

\author[Z.Yan et al.]{
Z. Yan, $^{1}$\thanks{E-mail:
yanza15@phas.ubc.ca}
A. J. Mead,$^{1,3}$
L. Van Waerbeke,$^{1}$
G. Hinshaw,$^{1}$
I. G. McCarthy$^{2}$
\\
$^{1}$Department of Physics and Astronomy, University of British Columbia, 6224 Agricultural Road, Vancouver, BC, V6T 1Z1, Canada\\
$^{2}$Astrophysics Research Institute, Liverpool John Moores University, 146 Brownlow Hill, Liverpool, L3 5RF, UK\\
$^{3}$Institut de Ci\`encies del Cosmos, Universitat de Barcelona, Mart\'i Franqu\`es 1, E-08028 Barcelona, Spain\\
}

\date{Accepted XXX. Received YYY; in original form ZZZ}

\pubyear{2019}
\hypersetup{draft}
\begin{document}
\pagerange{\pageref{firstpage}--\pageref{lastpage}}
\maketitle
\label{firstpage}

\begin{abstract}
We evaluate the ability of Convolutional Neural Networks (CNNs) to predict galaxy cluster masses in the BAHAMAS hydrodynamical simulations. We train four separate single-channel networks using: stellar mass, soft X-ray flux, bolometric X-ray flux, and the Compton $y$ parameter as observational tracers, respectively.  Our training set consists of $\sim$4800 synthetic cluster images generated from the simulation, while an additional $\sim$3200 images form a validation set and a test set, each with 1600 images. \ziangtxt{In order to mimic real observation, these images also contain uncorrelated structures located within 50 Mpc in front and behind clusters and seen in projection, as well as instrumental systematics including noise and smoothing.} In addition to CNNs for all the four observables, we also train a `multi-channel' CNN by combining the four observational tracers. The learning curves of all the five CNNs converge within 1000 epochs. The resulting predictions are especially precise for halo masses in the range $10^{13.25}M_{\odot}<M<10^{14.5}M_{\odot}$, where all five networks produce mean mass biases of order $\approx$1\% with a scatter of $\lesssim$20\%. The network trained with Compton $y$ parameter maps yields the most precise predictions. We interpret the network's behaviour using two diagnostic tests to determine which features are used to predict cluster mass. The CNN trained with stellar mass images detect galaxies (not surprisingly), while CNNs trained with gas-based tracers utilise the shape of the signal to estimate cluster mass.
\end{abstract}




\section{Introduction}
\label{sec:Introduction}

Galaxy groups and clusters are collections of several up to thousands of galaxies that are bound by their mutual gravity.  With masses in the range of $10^{13}$--$10^{15}$M$_{\odot}$, they are the most massive collapsed objects in the Universe. Their abundance, distribution, and morphology depends both on local physical processes and the underlying cosmological model. Stars typically comprise about 1\% of a cluster's mass (e.g., \citealt{leauthaud2011new, zu2015mapping}), while hot gas contributes anywhere from $\approx$7-13\% (depending on cluster mass; e.g., \citealt{allen2002,sun2009,pratt2009}), with the remainder residing in a dark matter halo.

The cluster mass function is a particularly sensitive probe of cosmological parameters and the evolutionary history of large-scale structure \citep[e.g. ][]{2005RvMP...77..207V, doi:10.1146/annurev-astro-081710-102514, 2016A&A...594A..24P}.  However, it is difficult to precisely and accurately measure cluster masses directly because they are dominated by dark matter.  Masses can be inferred from weak gravitational lensing data \citep[e.g.][]{2010arXiv1002.3952U, 2012ApJ...748...56S, hoekstra2015canadian, von2014weighing}, but the current signal-to-noise ratio of such observations limits the precision of individual cluster masses to typically (at least) tens of percent.  This is not sufficiently precise to be used directly for precision cosmology (via the mass function), but weak lensing is still a very important probe because it can be used to calibrate the mean bias\footnote{It has been shown from mock analyses of weak lensing observations of simulated clusters that weak lensing mass measurements yield a nearly unbiased mean mass estimate (e.g., \citealt{becker2011,bahe2012}).} of other tracers whose system-to-system scatter is lower.  Examples of such tracers include the total stellar mass or cluster richness, X-ray emission in the form of thermal bremsstrahlung and recombination lines from the hot intracluster medium (ICM), and the thermal Sunyaev-Zel'dovich effect (i.e., the inverse Compton scattering of cosmic microwave background photons off hot ICM electrons as they pass through clusters).  

Note that X-ray emission itself can be used to infer mass by combining spectroscopic measurements of the temperature profile with surface brightness measurements that strongly constrain the density profile, allowing one to infer a mass under the assumption of hydrostatic equilibrium.  How well this assumption holds is currently a subject of strong debate, with the level of deviation from hydrostatic equilibrium having been estimated to be anywhere from $40\%$ (i.e., the hydrostatic mass underestimates the true mass by this amount; e.g., \citealt{von2014weighing}) to only $\lesssim5\%$ (e.g., \citealt{melin2015,smith2016}). \ziangtxt{The Halo Occupation Distribution (HOD) \citep{peacock2000halo, seljak2000analytic} model links halo mass with galaxy properties. \citealt{moster2010constraints} studies relations between stellar mass and halo mass. The thermal Sunyaev-Zel'dovich (tSZ) effect is also known to be related to cluster masses through $Y_{500}-M_{500}$ relation where $Y_{500}$ is the Compton-$y$ paramster within $r_{500}$\citep{melin2011galaxy}. These studies link stellar mass, X-ray lumoinosity and tSZ with cluster mass, which suggests that they are important supplement to gravitational lensing to estimate cluster masses and probe cluster physics.}

Large-scale hydrodynamical simulations are playing an increasingly important role in calibrating the inference of cluster masses from observational data.  These simulations are now capable of capturing gravitational and gas dynamics on cosmological scales and can therefore provide large samples of realistic clusters in order to assess mass inferences statistically.  They also aid in understanding systematic effects that may hinder these inferences (see \citealt{borgani2011cosmological} for a review of hydrodynamical simulations).  However, even with these simulations, the complexities of substructure, morphology and small-scale physical processes, such as AGN feedback \citep{gitti2012evidence} and gas clumping \citep{nagai2011gas}, hinder the accuracy of many cluster mass estimates. \citet{yan2019analysis} used hydrodynamical simulations to assess the effect of cluster mis-centering on mass determinations.

Machine learning (ML) is a technique in which computer systems learn to analyze data without using explicit instructions or model parameterisations, but instead are `trained' to make decisions based on properties of the data itself. In astronomy, ML algorithms such as linear regression, decision tree, random forest and Principal Component Analysis have been widely used in model fitting and feature extraction (see \citet{baron2019machine} for a review). 

Artificial Neural Networks (ANNs) are a popular class of ML tools inspired by the way in which biological nervous systems, such as the brain, process information. ANNs use a hierarchy of simple functions, called activation functions, to construct a highly nonlinear function. Given their ability to mimic complicated functions, ANNs form the basis of many voice recognition and image identification tools.  A Convolutional Neural Network (CNN) is a category of ANN that is particularly useful in the field of object identification and image classification.  CNNs have been used by astronomers to classify galaxy morphology \citep{banerji2010galaxy}, to identify lensing shear \citep{lanusse2017cmu}, to generate cosmic webs \citep{rodriguez2018fast}, and to directly constrain cosmological parameters \citep{2019NatAs...3...93R}.

\citet{cohn2019multiwavelength, Armitage_2019} uses multiple machine-learning algorithms to estimate cluster mass from a set of observable quantities. \ziangtxt{\citet{green2019using} uses X-ray observational parameters, \citet{ntampaka2018deep} uses mock X-ray images, \citet{gupta2020mass} uses simulated microwave sky to train neural networks to predict cluster mass.} Here we extend their work and utilize CNNs to predict cluster masses from stellar mass data, X-ray data, Compton $y$ data, and from combinations of them.  The test data are the BAHAMAS hydrodynamical simulations \citep{McCarthy2017}.  We choose $M_{200}$ as the proxy for cluster mass.  This is the total mass within the characteristic radius $r_{200}$, the radius at which the cluster density falls to 200 times the critical density of the (simulated) universe.

The structure of this paper is as follows: \S\ref{sec:Method} describes the simulation data and the setup of our CNN; \S\ref{sec:Results} presents our results; \S\ref{sec:Interp} describes a test to understand the behaviour of the CNN; and \S\ref{sec:Conclusion} presents our conclusions.

\section{Data and Method}
\label{sec:Method}

\begin{figure}
    \centering
    \includegraphics[width=\columnwidth]{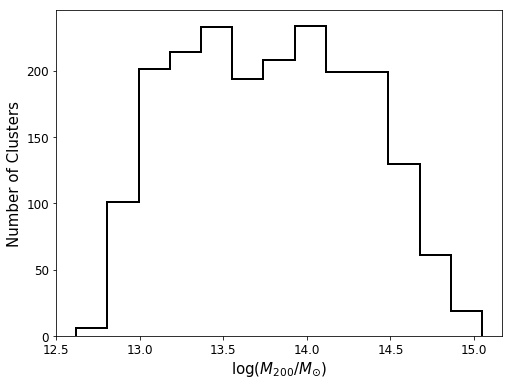}
    \caption{The mass distribution of galaxy clusters that we analyze from the BAHAMAS simulation.}
    \label{fig:mass_dist}
\end{figure}

\begin{figure*}
	\includegraphics[width=\textwidth] {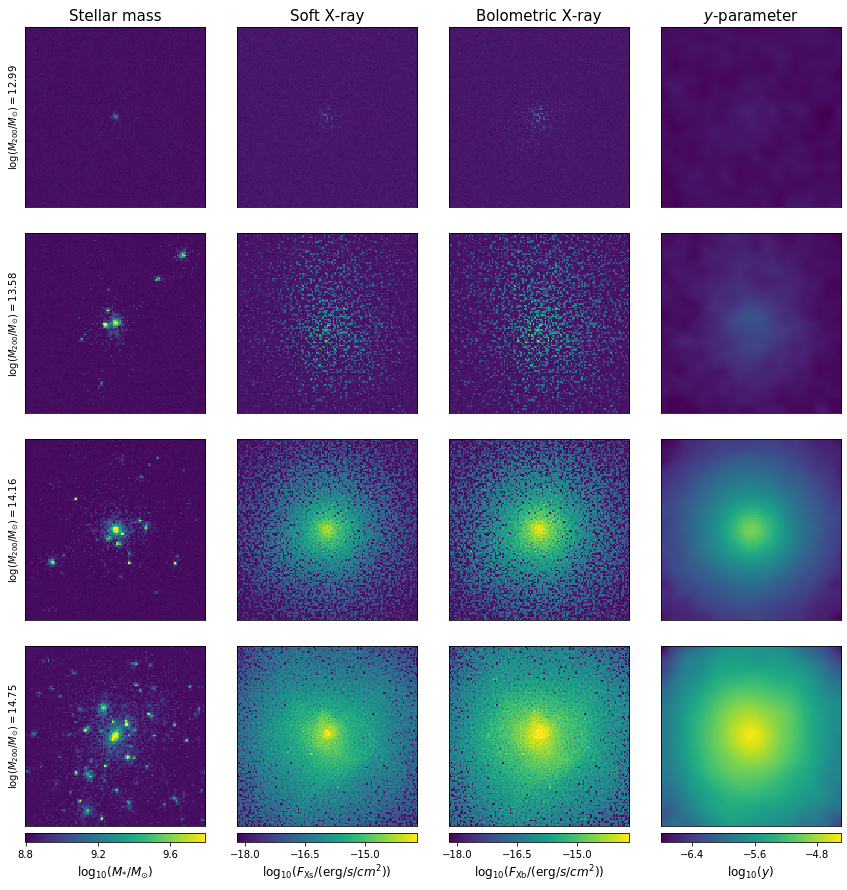}
    \caption{Selected cluster images from the BAHAMAS simulation. Each row is a cluster drawn from a different mass range, as indicated, and each column is a different observable: stellar density, soft X-ray luminosity, bolometric X-ray luminosity, and Compton $y$ parameter. The colour scales are the same across the mass range (rows). The angular size of all the image are 20 arcmin.}
    \label{fig:cluster_examples}
\end{figure*}
\subsection{The BAHAMAS Simulation}

We employ data from the BAHAMAS (BAryons and Haloes of MAssive Systems, \citealt{McCarthy2017, McCarthy2018}) simulations.  BAHAMAS is a suite of cosmological, hydrodynamical simulations run using a modified version of the TreePM SPH code {\tt GADGET3}. The simulations consist of $400~{\rm cMpc}/h$ periodic boxes containing $2 \times 1024^3$ particles (with equal numbers of dark matter and baryonic particles). The run we use adopts the WMAP 9-year best-fit cosmology with massless neutrinos \citep{Hinshaw2009}.

BAHAMAS includes subgrid treatments of important physical processes that cannot be directly resolved in the simulations, including metal-dependent radiative cooling, star formation, stellar evolution and mass-loss, black hole formation and growth, and stellar and active galactic nuclei (AGN) feedback. The subgrid models were developed as part of the OWLS project \citep{Schaye2010}. The parameters governing the efficiencies of AGN and stellar feedback were adjusted so that the simulations approximately reproduce the observed galaxy stellar mass function for ${\rm M_*} \geq 10^{10}~{\rm M_{\odot}}$ and the hot gas fraction--halo mass relation of groups and clusters, as determined from high-resolution X-ray observations of local systems.  As shown in \citet{McCarthy2017}, the simulations match the galaxy--halo--tSZ--X-ray scaling relations of galaxies and groups and clusters.

For the present study, friends-of-friends (FoF) halos are selected from the dark-matter-only simulation that accompanies the BAHAMAS hydrodynamical simulations\footnote{We select halos from a dark matter-only simulation so as to facilitate comparisons with hydrodynamical runs that vary feedback and the cosmological model.}.  We select up to 200 halos in each of 10 mass bins of width of $0.25$ dex, spanning the range $M_{200} = 10^{13} - 10^{15} ~{\rm M_{\odot}}$, resulting in a sample of almost 2000 halos (some bins have slightly fewer than 200 halos).  This sample is then matched to the BAHAMAS hydrodynamical simulation.  The resulting number distribution of clusters as a function of mass is shown in Fig.~\ref{fig:mass_dist}. The  distribution is not perfectly flat because the hydrodynamical masses are different from the underlying dark-matter-only masses.

We tag all particles (gas, dark matter, and stellar) within $2 r_{200}$ of the most bound particle (MBP) for analysis.  We also generate a catalog of simulated galaxies within this radius that have $M_{\rm{gal}} > 10^{10} ~\rm{M_{\odot}}$.  (Simulated galaxies are defined as the stellar component of self-gravitating substructures identified with the {\tt SUBFIND} algorithm.)

The soft and bolometric X-ray luminosity of each gas particle is provided with the simulation.  For the tSZ signal, a quantity $\Upsilon$ is calculated for each gas particle \citep{McCarthy2018},
\begin{equation}
\Upsilon \equiv \sigma_{T} \frac{k_{b} T}{m_{e} c^{2}} \frac{m}{\mu_{e} m_{H}},
\end{equation}
where $T$ is the gas particle's temperature, $m$ is the gas particle's mass, $\mu_e$ is the mean molecular weight per free electron of each gas particle, and $m_H$ is the atomic mass of hydrogen.

\subsection{Data set and image generation}

The data sets used to train the neural networks are images of each of the four observables derived from the simulated cluster sample. The simulated clusters are provided at redshift 0, but we place them at random redshifts between 0.03 and 0.07 (with a uniform distribution) when we produce images. The cluster catalog contains $\sim$2000 clusters, but this is insufficient to train the neural network to the desired level of precision. \ziangtxt{To overcome this, we generate four images of each cluster by projecting along four different directions: $x$, $y$, $z$-axis and along $(\sqrt{2}/2, -\sqrt{2}/2, 0)$. To make the images of same cluster look more different, we rotate it with a random azimuthal angle before projecting. To properly include the correlated structure and foreground contamination which are difficult to remove in real observation, we also project all the particles within 50 Mpc in front and 50 Mpc behind each cluster to the images. To make the images of the same cluster look more different from each other, all the clusters are rotated with a random angle around the line of sight before projection.} In the end, we have $\sim$ 8000 clusters at different redshifts with which to train the neural networks.  

The image of each cluster is made by projecting it onto the $x$-$y$ plane and binning the particles onto an $120\times120$ grid with an overall angular size of 20 arcmin. For a cluster at redshift $z$, the signal in pixel $(i, j)$ for each observable is obtained as follows.

{\bf Stellar density} - We evaluate the stellar surface density in each pixel as
\begin{equation}
I_{ij} = \sum_{p \in (i, j)} M_{\rm s} (\boldsymbol{r}_p)/S,
\end{equation}
where the sum is over all stellar particles that project into pixel $(i,j)$, $M_{\rm s}(\boldsymbol{r}_p)$ is the stellar mass of particle $p$ (located at position $\boldsymbol{r}_p$ with respect to the cluster centre), and $S$ is the physical area of pixel $(i,j)$. The angular coordinates of pixel $(i,j)$ are
\begin{equation}
\boldsymbol{\theta}_p = (x_p, y_p) / d_{\rm A}(z),
\end{equation}
where $(x_p, y_p)$ are the $x$ and $y$ components of $\boldsymbol{r}_p$ and $d_{\rm A}(z)$ is the angular diameter distance to redsfhit $z$.

{\bf X-ray emission} - We convert luminosity into flux using $F = L/4\pi d_{\rm L}(z)$, where $d_{\rm L}(z)$ is the luminosity distance to redshift $z$. The signal in pixel $(i,j)$ is the flux due to all gas particles that project into that pixel,
\begin{equation}
I_{ij} = \sum_{p \in (i, j)} F(\boldsymbol{r}_p) = \sum_{p \in (i, j)} \frac{L(\boldsymbol{r}_p)}{4\pi \, d_{\rm L}(z)}.
\end{equation}

{\bf Compton $y$ parameter} - The signal in pixel $(i,j)$ is obtained by summing $\Upsilon/S$ \citep{McCarthy2018} over all gas particles that project into that pixel,
\begin{equation}
I_{ij} = \sum_{p \in (i, j)} \Upsilon(\boldsymbol{r}_p)/S.
\end{equation}

For low-mass clusters (those with $2\theta_{200} < 20'$), all cluster particles reside within the image, while for high-mass clusters, some particles extend outside the image and are lost.  Our choice of image size strikes a balance between performance and computation time. 

In order to mimic realistic data, we add noise to our images and smooth them to mimic a telescope point spread function (PSF).  For stellar images, we take the {\em rms} to be 1/10 the mean mass across the whole sample giving a signal-to-noise-ratio roughly 10, which mimics an SDSS-like observation\citep{abazajian2009seventh}. No smoothing is applied since most optical telescopes have a beam size smaller than our pixel size.  The gas-based images have Gaussian random noise added and are then smoothed with a Gaussian beam.  For the X-ray images, the {\em rms} noise and beam size are chosen to match the Chandra HRI sensitivity and FWHM, respectively \citep{abazajian2009seventh}.  For the Compton $y$ image, the {\em rms} noise is taken to be $10^{-8}$ per pixel and the beam FWHM is $1.4$ arcmin, corresponding to an {\sl ACT}-like experiment \citep{hasselfield2013atacama}. We have also considered a {\sl Planck}-like experiment with {\em rms} noise $10^{-6}$ and a FWHM of $9.66$ arcmin \citep{aghanim2016planck}, but the CNN to performed quite poorly in this case. The parameters discussed above are summarized in Table \ref{table:table_datasets_summary}. Images of each observable, in four clusters selected to have different masses, are shown in Fig.~\ref{fig:cluster_examples}. 

\begin{table}
\begin{tabular}{lllll}
\hline\hline
Signal         & Label & Units       & Noise         & Beam \\ 
    &  & & [{\em rms}]   & [FWHM] \\ 
\hline
Stellar Mass     & Star          & ${\rm M}_{\odot}$ & $2.14\times 10^{11}$ & -  \\
Soft X-ray       & Fxs           & erg/s/cm$^2$       & $9\times 10^{-16}$ & 4$^{\prime\prime}$ \\
Bolometric X-ray & Fxb           & erg/s/cm$^2$      & $9\times 10^{-16}$ & 4$^{\prime\prime}$ \\
Compton $y$    & Ypar          & -        & $10^{-8}$           & 1.4$^{\prime}$  \\
\hline\hline
\end{tabular}
\caption{Simulated data set properties. The labels are used throughout this paper.}
\label{table:table_datasets_summary}
\end{table}


\begin{figure*}
\subfloat[Single-channel]{%
  \includegraphics[clip,width=\textwidth]{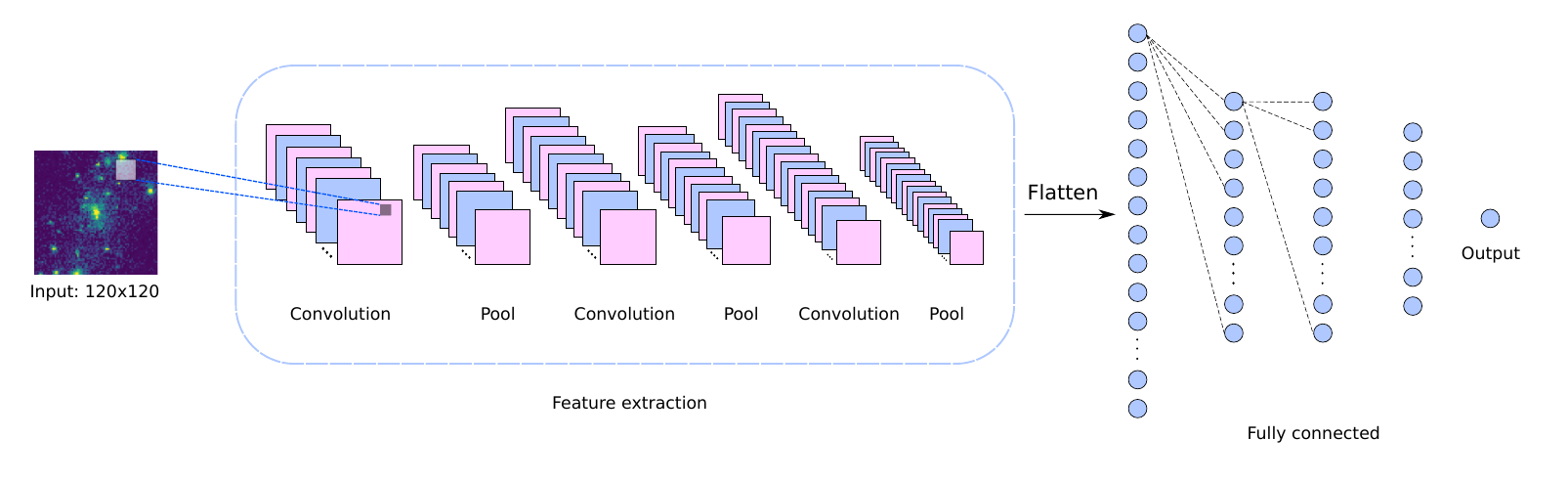}
  \label{fig:cnn_sc}%
}

\subfloat[Multi-channel]{%
  \includegraphics[clip,width=\textwidth]{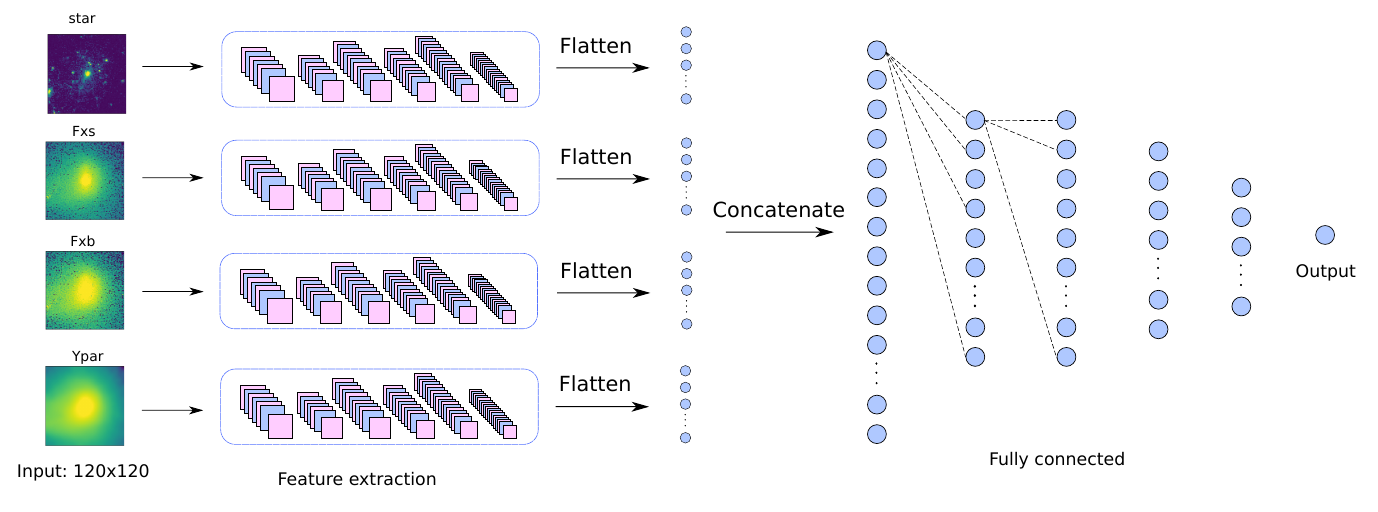}
  \label{fig:cnn_mc}%
}

\caption{Upper panel: Architecture of the single-channel CNN used in this analysis. Our network utilizes three convolutional and pooling layers for feature extraction and four fully connected layers for parameter estimation. Lower panel: Architecture of the multi-channel CNN.  The four channels take images: Star, Fxs, Fxb, and Ypar, respectively and perform feature extraction independently. The feature extraction layers have the same structure as the single-channel portion outlined in the upper panel.}
\end{figure*}

\subsection{Artificial Neural Network}

An Artificial Neural Network (ANN) is a function which maps inputs to outputs,
\begin{equation}
{\tt ANN}\left(I\right) = O,
\end{equation}
where $I$ is the input and $O$ is the output. In practice, the inputs can be images, sounds, text, etc., and the output can be a parameter to measure, a classification, and so on.  A typical ANN is a sequential nest of functions that are defined on each layer of the network. In the simplest case of a feed-forward neural network, the neuron layers are evaluated in sequence, passing information from layer to layer.  The output, $a_k^l$, of the $k$-th neuron in the $l$-th layer may be written as
\begin{equation}
a_k^l = f \left(\sum_{j} W^l_{jk} a_j^{l-1} + b_k^l\right),
\end{equation}

where $f$ is called an activation function (our choice of $f$ is defined in the following section), $W^l_{jk}$ is a matrix of weights, and $b_k^l$ is a vector of additive biases.  $a_k^0$ is the input data, $I$, and $a$ in the last layer is the output, $O$. Schematically, $W^l_{jk}$ connects the $j$-th neuron in layer $(l-1)$ to the $k$-th neuron in layer $l$. The number of neurons and layers, or equivalently, the dimensions of $W^l_{jk}$ and $b_k^l$ are called the architecture of the ANN. Given the architecture and activation functions, the ANN is completely specified $W^l_{jk}$ and $b_k^l$.

ANN training is a fitting procedure to determine the parameters $W$ and $b$ required to reproduce known information (so-called 'labels') from data.  The labels can be categories (for a classification task) or quantities (for a \ziangtxt{regression task}), and so on. For example, an ANN designed to recognise hand-written numbers is a classifier that takes hand-written images of numbers as input, and generates numbers as output labels.

ANN training involves iteratively optimizing the weights so as to minimize the difference between the output labels and the known labels, as quantified by the loss function.  The ANN is initialized with random weights and biases, then, during each iteration ('epoch'), the training data is provided to the ANN and outputs are predicted from them.  The weights and biases are updated to reduce the loss function by an algorithm called an optimizer. The training is complete when the loss function converges. \ziangtxt{In order to validate the model, a `validation set' (whose labels are also known) is needed. The loss function is calculated on the validation set during each epoch to monitor the training. The training is considered to be finished when the validation loss converges. After that, a 'test set' is then supplied to the ANN.}  If the ANN gives accurate predictions for the test set, then one can safely use it to predict labels from data whose labels are not known.

\subsection{Convolutional Neural Network}

In our analysis we use a category of ANN called a Convolutional Neural Network (CNN)\ziangtxt{(see, for example, \citealt{8286426} for a review on CNN)}. The typical input of a CNN is a two-dimensional image, and the CNN uses convolution layers to extract features from them (for example, textures, edges, gradual changes and so on). Unlike fully-connected layers, in which each neuron is connected to each of the previous neurons, convolutional layers pass forward information from a small neighborhood around each neuron. \ziangtxt{A convolution layer is comprised of several filters, which are smaller than the input image. The filtered image is given by}
\begin{equation}
\label{eq:cnn_filter}
I^F_{ij} = \sum_{i'j'} F_{i'j'} I_{i+i',j+j'},
\end{equation}
\ziangtxt{where the sum runs over the filter pixels, centred on $(i',j')=(0,0)$.}
The output is called a feature image, and within the same convolutional layer, different filters extract different kinds of features (for example, horizontal and vertical textures). The parameters in a filter define a set of weights that are optimized during training.  The feature images are downsized into a `pooling layer', so the feature images gets smaller as they pass through convolution-pooling layers. Different convolution layers can be designed to extract information on different scales by tuning the filter size, and the feature image size. \ziangtxt{In our analysis, we take $3\times3$ square filters; which means in \eqref{eq:cnn_filter}, $i'$ and $j'$ take the value $\{-1, 0, 1\}$.} The pooling filter is $2\times 2$ with a stride of 2 pixels. This means that each pixel in the pooling layer is the average of a $2\times 2$ patch in the previous feature image with a stride of 2 pixels. This process is called 'average pooling', which downsizes the feature image by a factor of 2. As the feature images are downsized from layer to layer, the deeper convolution layers extract larger scale features. By using convolution and pooling layers, one can also reduce the computational cost and make the result easier to interpret. A sequence of convolution-pooling layers is flattened into a one-dimensional layer followed by fully connected layers to further parameterize the features.

In our application, we utilize a CNN to predict cluster masses, so the output layer is a single neuron: the cluster mass. In the training set, we label each cluster with the $M_{200}$ value calculated by summing the masses of all simulated particles within $r_{200}$ of the cluster centre. In the rest of this paper, we denote this value as $M_{\rm true}$, and we denote the CNN-predicted mass by $M_{\rm pred}$.  \ziangtxt{For each training run, we randomly select images of $\sim 4800 (60\%)$ simulated clusters as the training set, $\sim 1600 (20\%)$ as the validation set and the remaining $\sim 1600 (20\%)$ as the test set. The training and testing sets are carefully split so that we never train on a simulated cluster and then test on the same cluster as viewed from a different angle.}

We train the four `single-channel' CNNs with the four data sets described in Table~\ref{table:table_datasets_summary}. We can write
\begin{equation}
{\tt CNN}^{c}\left(I^{c}_{ij}\right) = M_{\rm pred}
\end{equation}
where $I^{c}_{ij}$ is the image of tracer $c\in$\{Star, Fxs, Fxb, Ypar\}, $(i,j)$ is the 2-D pixel index, and $M_{\rm pred}$ is the predicted $M_{200}$.  To assess the advantage of multiple tracers, we also train a `multi-channel' CNN, denoted ${\tt CNN}^{mc}$, by simultaneously feeding all four data sets into one neural network,
\begin{equation}
{\tt CNN}^{mc}\left(I_{ij}^{\rm Star}, I_{ij}^{\rm Fxs}, I_{ij}^{\rm Fxb}, I_{ij}^{\rm Ypar}\right) = M_{\rm pred}.
\end{equation}

For each layer except the output layer, we use the Rectified Linear Unit (ReLu) as our activation function \ziangtxt{\citep{nair2010proceedings}}.  This is defined as $f(x)\equiv \max \{0, x\}$. To prevent over-fitting, we force a 20\% dropout between fully-connected layers \ziangtxt{\citep{srivastava2014dropout}}, which means that, for each training epoch, 20\% of the weights (randomly selected) between those layers are set to zero.  A  dropout fraction slows down the training, so we chose this value to prevent over-fitting while keep the training fairly fast. For the output layer, we use the mean-squared-logarithmic-error as our loss function, defined as
\begin{equation}
\delta \equiv \left\langle\left(\log{M_{\rm pred}} - \log{M_{\rm true}}\right)^2 \right\rangle.
\label{eq:loss_function}
\end{equation}
During training, the loss function of the validation set (the so-called validation loss) is calculated in each epoch to monitor the progress of the training. Our convergence criterion is discussed below.

Our CNN is implemented using the {\tt keras} package with a {\tt Tensorflow} back-end written in {\tt python}. Our network architecture is similar to that used by \citet{ntampaka2018deep}, which is a simplified version of that used by \citet{simonyan2014very}:
\begin{enumerate}[1.]
\item $3\times3$ convolution with 16 filters
\item $2\times2$, stride-2 average pooling
\item $3\times3$ convolution with 32 filters
\item $2\times2$, stride-2 average pooling
\item $3\times3$ convolution with 64 filters
\item $2\times2$, stride-2 average pooling
\item Flatten
\item Fully-connected with 200 neurons
\item 10\% dropout
\item Fully-connected with 100 neurons
\item 10\% dropout
\item Fully-connected with 100 neurons
\item 10\% dropout
\item Fully-connected with 20 neurons
\item Output neuron
\end{enumerate}

For our multi-channel network, each data channel is convolved, pooled, and flattened separately, using the same architecture as the single-channel network. As shown in Fig.~\ref{fig:cnn_mc}, the flattened layers from each channel are concatenated into one flattened layer, followed by fully-connected layers with the same architecture as the single-channel network.

We use {\tt RMSprop} \citep{hinton2012overview} as our optimizer because it converges quickly in this application. We set the learning rate (the step size in the parameter space) to be 0.01 with decay rate of $10^{-4}$. We tested other combinations of optimizers and learning rates, but this choice gave the best performance.  The training data is divided into batches of 50 images each. In one training epoch, the network is trained through each batch separately, and the CNN weights are obtained by averaging over all batches.  Each network was trained for 1000 epochs on 2 GPUs with 6 CPUs.  The training took about 20 minutes for a single-channel network, and 45 minutes for the multi channel network.

\begin{figure}
    \centering
    \includegraphics[width=\columnwidth]{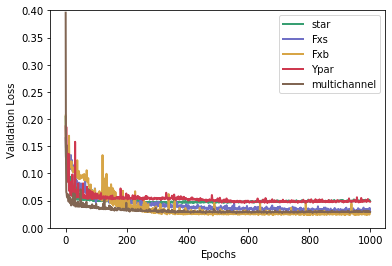}
    \caption{The learning curves for each of our five CNNs as a function of training epoch. The $y$-axis is the loss function on the validation set defined in equation~\eqref{eq:loss_function}.}
    \label{fig:learning_curve}
\end{figure}

Fig.~\ref{fig:learning_curve} shows the `learning curve' (validation loss as a function of training epoch) for each of our CNNs. During training, the validation loss drops quickly at first, then converges after $\sim$600 epochs.  The final CNN weights are taken to be those which gave the minimum validation loss during training.

{}

\section{Results}
\label{sec:Results}

\begin{figure*}
    \centering
    \includegraphics[width=2.\columnwidth]{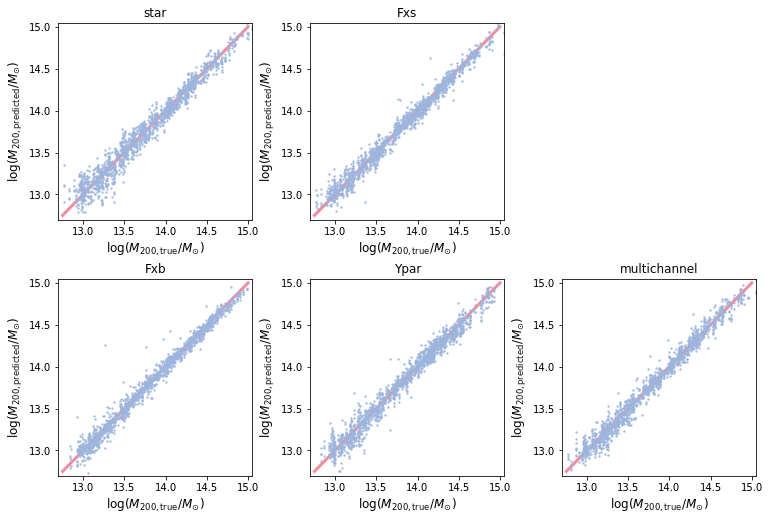}
    \caption{The CNN-predicted cluster mass vs. the true mass for each cluster in the test set.  Each tracer is shown separately, for clarity.}
    \label{fig:mass_to_mass}
\end{figure*}{}

\begin{figure}
    \centering
    \includegraphics[width=\columnwidth]{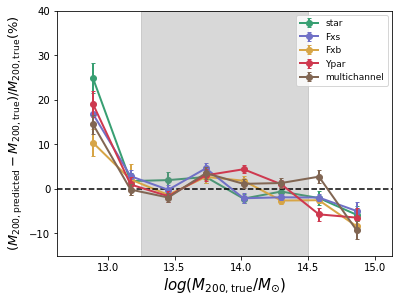}
    \caption{The bias in our CNN-predicted cluster masses as a function of the true mass, for each tracer.  The plotted uncertainties show the standard deviation of the bias in each bin.  The bias is small within the central mass range of $13.25 < \log(M_{\rm 200, true}/M_{\odot}) < 14.5$ (the shaded region).}
    \label{fig:mass_bias_mass}
\end{figure}{}

\begin{figure}
    \centering
    \includegraphics[width=\columnwidth]{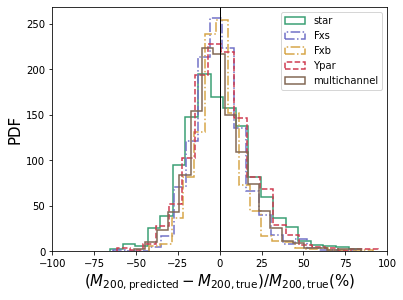}
    \caption{The probability distribution of the mass bias for each tracer, for mass bins in the range $13.25 < \log(M_{\rm 200, true}/M_{\odot}) < 14.5$ (the shaded region in Fig.~\ref{fig:mass_bias_mass}).}
    \label{fig:mass_bias_hist}
\end{figure}{}

\begin{table}
\centering
\begin{tabular}{lrrr}
\hline
\hline
Dataset  & $\left\langle\log{\frac{M_{\textrm{predict}}}{M_{\textrm{true}}}}\right\rangle$ & $\left\langle\frac{M_{\textrm{predict}}}{M_{\textrm{true}}}\right\rangle-1(\%)$ & $\left\langle \mathrm{RMS} \right\rangle$ \\
\hline
star & -0.01$\pm$0.003 & -0.516$\pm$0.621 & 19.028\\
Fxs & -0.007$\pm$0.002 & -0.349$\pm$0.517 & 16.49\\
Fxb & -0.004$\pm$0.002 & 0.094$\pm$0.524 & 16.036\\
Ypar & 0.002$\pm$0.002 & 1.814$\pm$0.559 & 17.662\\
multichannel & -0.001$\pm$0.002 & 1.075$\pm$0.575 & 17.693\\
\hline
\hline
\end{tabular}
\caption{The mean mass bias ($\Delta M \equiv M_{\rm pred} - M_{\rm true}$) and scatter obtained from the test set for $13.25 < \log(M_{\rm 200, true}/M_{\odot}) < 14.5$.}
\label{table:results}
\end{table}

Our cluster mass predictions are shown in Fig.~\ref{fig:mass_to_mass}.  For each cluster in the test set, we show the CNN-predicted mass versus the true $M_{200}$ measured in the simulation.  In this rendition, all five data sets produce similar results.  Fig.~\ref{fig:mass_bias_mass} shows the fractional mass bias for each tracer as a function of the true mass. \ziangtxt{From Fig.~\ref{fig:mass_bias_mass}, we see a clear tendency that the CNN generally {\em over}-predicts the mass by $\sim$20\% in the lowest mass bin, while it {\em under}-predicts the mass by $\sim$10\% in the highest mass bin. The is due to the fact that for these extreme masses there are not enough samples, so the CNNs tend to predict towards the mean mass of the whole sample. To mitigate the this ‘towards-the-mean’ bias, one needs to extend the mass range of training set than the test set, or alternatively, only trust the results of test clusters with masses close to the mean.}

Within the central mass range of $13.25 < \log(M_{\rm 200, true}/M_{\odot}) < 14.5$ (the shaded region in Fig.~\ref{fig:mass_bias_mass}), the mass bias is quite small. Histograms of the mass bias in these central bins are shown in Fig.~\ref{fig:mass_bias_hist}.  Each tracer is plotted as a separate colour, with the gas-based tracers plotted as dashed curves, for clarity.  A summary of our numerical results, both the average bias and the {\em rms} scatter, are given in Table~\ref{table:results}.

The average mass bias, $\Delta M/M_{\rm true}$, in the central mass bins is on the order of 1\% with an uncertainty of $\sim$0.5\%. The uncertainty per individual cluster is of order 15\% (Table~\ref{table:results}).  Somewhat surprisingly, the multi-channel network is not the most precise.  We assume this is due to limitations in the CNN architecture to synthesize information across all 4 tracers. 

\ziangtxt{\citet{Armitage_2019} applies a machine learning method on cluster masses and reports a $7\%$ mass scatter. However, they use multiple observables derived from simulations to train their model. These observables may suffer from uncertainty in real observation. In addition, they do not include observational effects such as instrument noise or beams, which will degrade the performance of the mass estimation. \citet{henson2016impact} evaluate the performance of conventional mass estimation techniques applied to the BAHAMAS hydrodynamical simulations.  They fit azimuthally averaged shear profile of each simulated cluster with both NFW and Einasto models with the cluster mass as a free parameter. By comparing best-fit mass with the true cluster mass, they find mass biases of $\Delta M/M_{\rm true} = -8.9^{+0.3}_{-0.2}$\%, and $-6.4^{+0.3}_{-0.2}$\% for the NFW and Einasto profile respectively. \citet{yan2019analysis} analyze the same BAHAMAS catalog by fitting an NFW model to the density profile of all particles in a cluster and found a mean mass bias of $-10$\%. These studies are based on weak lensing profiles which is an unbiased tracer of the cluster masses, while the present study uses biased tracers like galaxy or gas. Moreover, the previous studies do not include observational effects such as noise and smoothing. We conclude that our CNN-based results are more accurate than these profile-based analyses performed on the same hydrodynamical simulation even with biased tracers, possibly due to limitations in the profile models they use. As we will see in the next section, CNN is capable of extracting shapes, orientations and substructures from 2-D cluster images, which contains more information about the cluster masses.}

As a reference to real observation, \citet{zhang2008locuss} use scale relations of X-ray observations to evaluate real cluster masses and get an individual mass uncertainty of $\sim 30\%$; \citet{bleem2015galaxy} also use scale relation of tSZ signal to estimate mass for SPT galaxy clusters and get a mass uncertainty of $\sim 24\%$ for each cluster. \citet{hoekstra2015canadian} use weak lensing techniques to evaluate the masses of clusters.  They estimate an uncertainty of about 20\% which, if correct, indicates that the precision of our CNN-based method is not significantly better than weak lensing analysis. However, the weak lensing analysis is generally performed on more massive clusters which are not readily available in our simulation, so the comparison is not completely apt.

\ziangtxt{There are 3 causes of the mass bias of our CNN prediction: 1) the underlying scatter between cluster mass and morphology correlation; 2) the observational biases caused by smoothing and noise; 3) the imperfection of our CNN algorithm. The first cause is not able to overcome by observation; the second one could be suppressed by carefully handling instrument systematics; the third one could be suppressed by improving the CNN architecture and training setup. In addition, to apply this method on real observation, one needs to take care of the difference between simulation and observation. This can be done by either comparing simulated galaxy clusters with real clusters, or introduce real data in the testing set, which are left for future work.}

\ziangtxt{In order to evaluate the impact of the foreground and background interlopers, we also train a set of CNNs with images which do not have fore- and background. The scatter of mass bias is lower than our fiducial results by $\sim 2\%$. This indicates the presence of uncorrelated structure along the line-of-sight has only a marginal impact on our results.}

\section{Interpreting the CNN Performance}
\label{sec:Interp}

\begin{figure*}
\centering
  \includegraphics[width=0.8\textwidth]{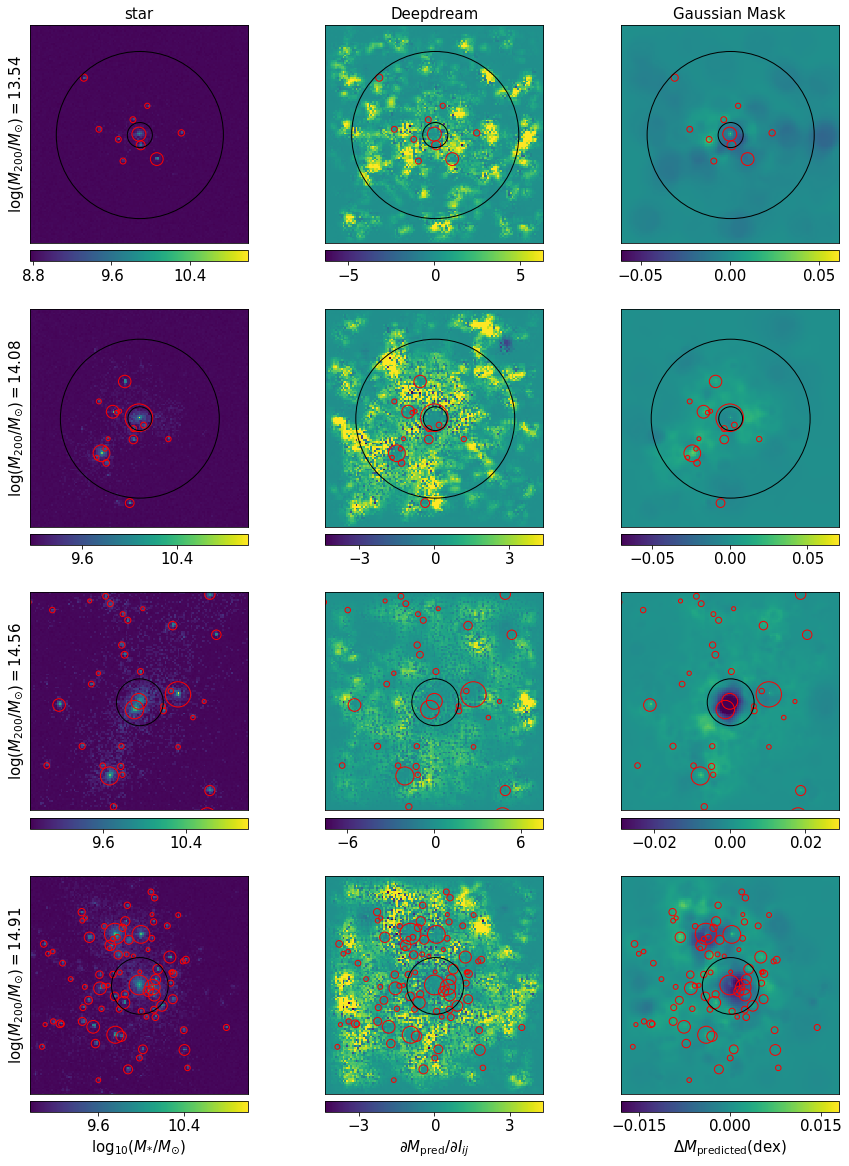}
\caption{{\it Left column}: stellar mass images of 4 galaxy clusters selected to cover our mass range; {\it middle column}: the relative signal change, $\propto \Delta M_{\rm pred}$, after two Deep Dream iterations; {\it right column}: the signal change, $\Delta M_{\rm pred}$, when masking the image with a Gaussian mask centered, in turn, on each image pixel (see text for details). The inner black circles indicate $0.15 R_{200}$, while the outer circles indicate $R_{200}$. The red circles highlight galaxy positions, with radii that are proportional to the galaxy's mass.}
\label{fig:star_DD_grad}%
\end{figure*}

\begin{figure*}
\centering
  \includegraphics[width=0.8\textwidth]{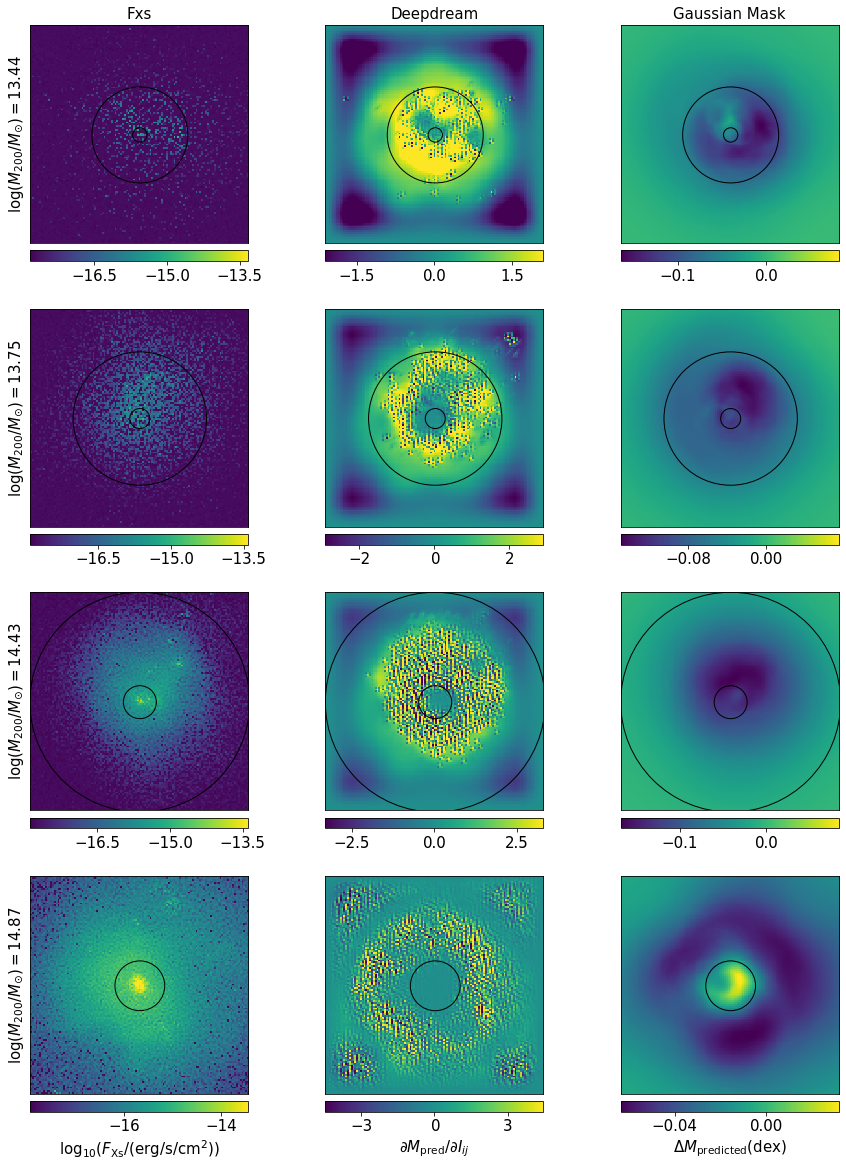}
\caption{Left column: soft X-ray images of 4 galaxy clusters; middle column: change of signal after two Deep Dream iteration; right column: change of mass prediction when masking the image with Gaussian masks centered at each pixels.The inner circles shows the radius of $0.15R_{200}$ and the outer circles $R_{200}$}
\label{fig:Fxs_DD_grad}%
\end{figure*}

\begin{figure*}
\centering
  \includegraphics[width=0.8\textwidth]{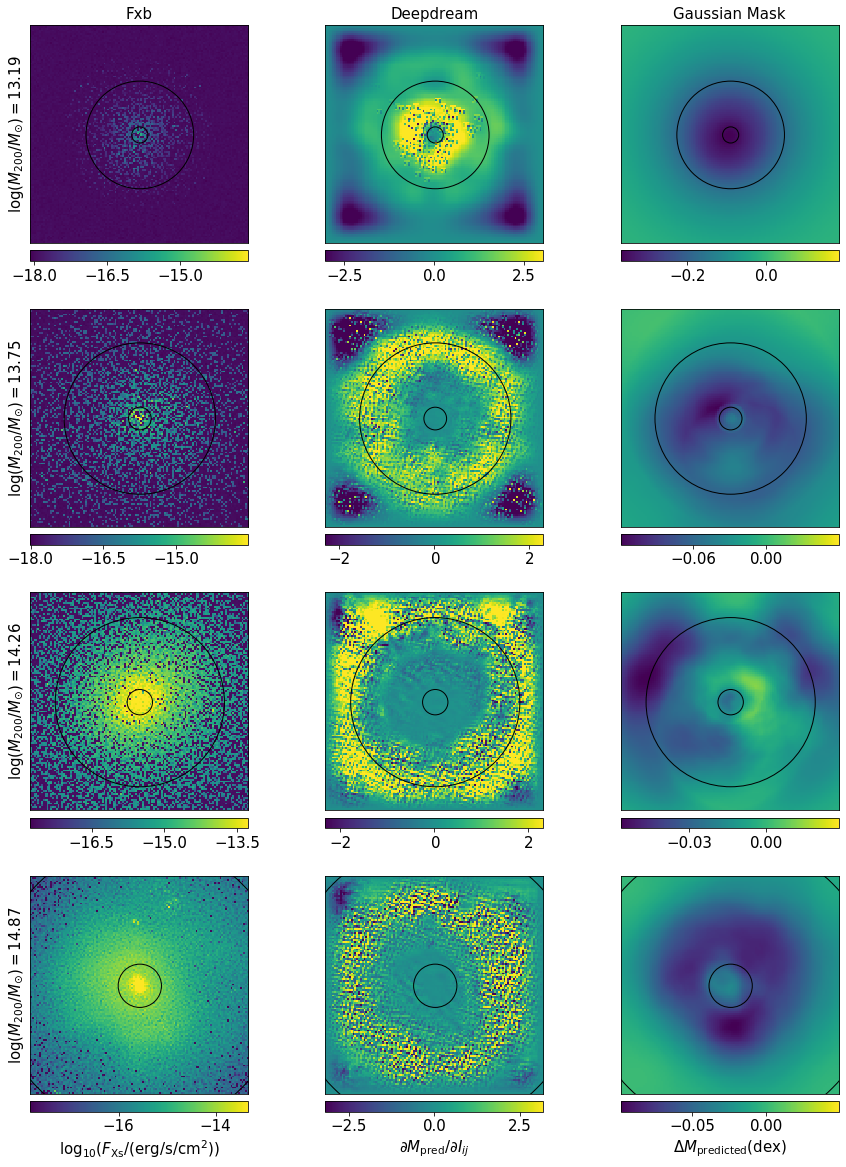}
\caption{Left column: bolometric X-ray images of 4 galaxy clusters; middle column: change of signal after two Deep Dream iteration; right column: change of mass prediction when masking the image with Gaussian masks centered at each pixels.The inner circles shows the radius of $0.15R_{200}$ and the outer circles $R_{200}$}
\label{fig:Fxb_DD_grad}%
\end{figure*}

\begin{figure*}
\centering
  \includegraphics[width=0.8\textwidth]{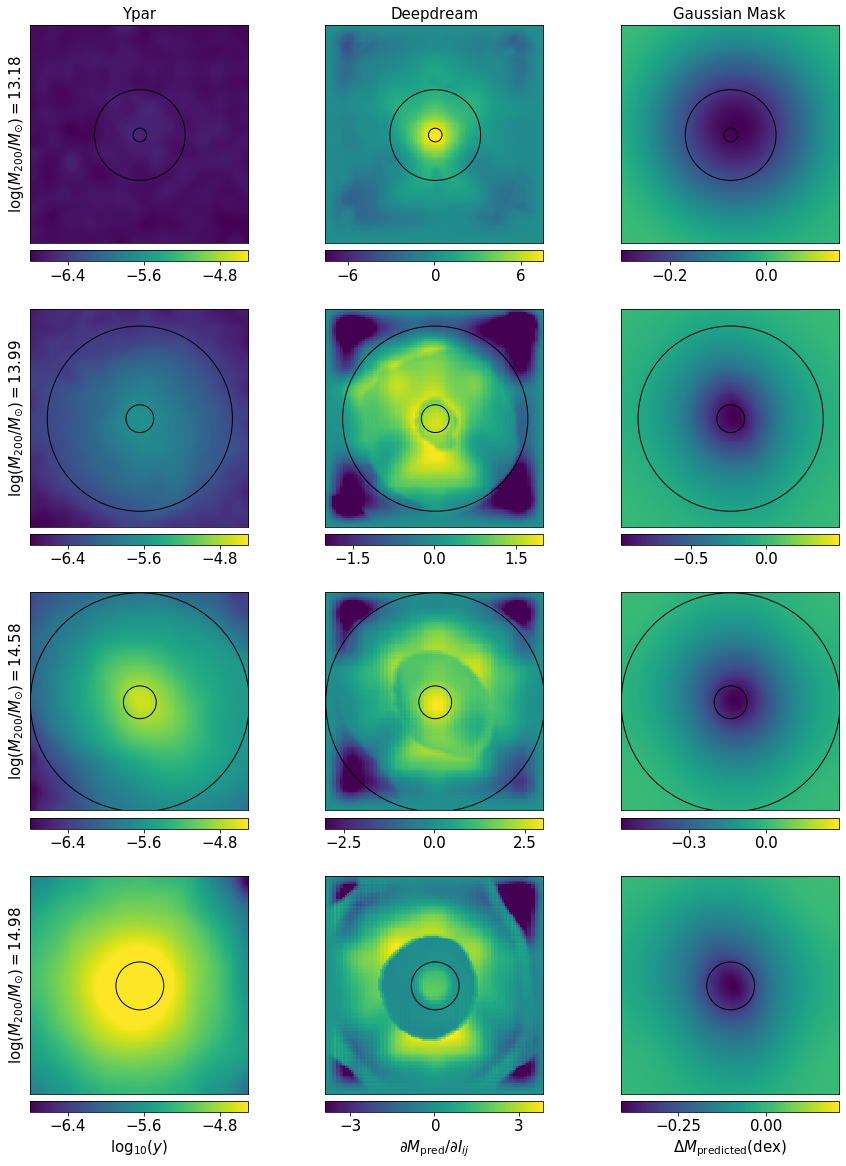}
\caption{Left column: $y$ parameter images of 4 galaxy clusters; middle column: change of signal after two Deep Dream iteration; right column: change of mass prediction when masking the image with Gaussian masks centered at each pixels.The inner circles shows the radius of $0.15R_{200}$ and the outer circles $R_{200}$}
\label{fig:ypar_DD_grad}%
\end{figure*}

Cluster masses are traditionally estimated using scaling relations based on known physics.  For example, in relaxed clusters, X-ray luminosity is related to cluster mass via the virial theorem.  In contrast, neural networks contain a large number of parameters, which makes their behaviour difficult to interpret.  What makes the network predict a particular value of mass?  What cluster feature(s) is it sensitive to?  In this section, we attempt to interpret our single-channel networks in two ways.

{\bf Deep Dream:} Google's Deep Dream (DD) \citep{mordvintsev2015deepdream} is an iterative, gradient ascent algorithm that is applied to an input image to determine which image pixels affect a particular output neuron the most.  In our application, we have one output neuron, $M_{\rm pred}$, so DD may be expressed in the form
\begin{equation}
    I^{(p)}_{ij} = \left. I^{(p-1)}_{ij} + \alpha\,\frac{\partial M_{\rm pred}}{\partial I_{ij}}\right|_{I^{(p-1)}},
\end{equation}
where $I^{(p)}_{ij}$ is the image at the $p$-th iteration of the algorithm, and $\alpha$ is the step size.  For small $\alpha$, the difference between successive image iterations is proportional to the gradient of $M_{\rm pred}$ with respect to the image.

We ran one iteration of DD for each data tracer, selecting images from a range of true cluster masses.  (We also ran 2 iterations, as did \citet{ntampaka2018deep}, and found similar results.)  The gradient images for these examples are shown in the middle columns of~\Crefrange{fig:star_DD_grad}{fig:ypar_DD_grad}.  In the stellar mass examples, the pixels that affect the predicted mass the most (rendered in yellow) appear to lie mostly {\em adjacent} to the galaxy locations.  This suggests that ${\tt CNN}^{\rm star}$ is mainly triggering on the number and size of galaxies in the image.

The gas-based tracers are more diffuse and symmetric, and this is reflected in the DD gradient images.  For the X-ray tracers, the gradient images images are quite granular, reflecting the granularity of the input images. But the critical information captured by the CNNs appears to be the shape of a cluster. For example, in the third row of Fig.~\ref{fig:Fxs_DD_grad}, we see that the Fxs image has substructure at the top-right, which is also seen in the gradient image.  The central regions appear to be relatively uninformative, in agreement with the conclusions of \citet{ntampaka2018deep}. \ziangtxt{In addition, the lack of sensitivity of the central region with the DD images generally mimics the shape of the cluster itself (it is clear, for instanc, in the lowest two rows of Fig.~\ref{fig:Fxb_DD_grad}).}  For the Compton $y$ images, the DD gradient images show two contours at different radius for massive clusters, both sketching the outskirt of the cluster without the fine granularity seen in the X-ray tracers.

{\bf Gaussian Mask:} A somewhat complimentary approach to interpreting the CNN performance is to examine the predicted mass when selected regions of the image are masked.  For this study, we define a Gaussian mask in the image plane as
\begin{equation}
    {\rm Mask}_{ij} = 1-\exp\left[-\frac{\left(i-a\right)^2+\left(j-b\right)^2}{2\sigma^2}\right],
\end{equation}
where $a$ and $b$ define the centre of the mask in pixel coordinates, and we take $\sigma = 5$ pixels, corresponding to 1.25 arcmin for our images.  For each $a$ and $b$ in the image plane, we multiply the original image by this mask, then use the pre-trained CNN to (re)predict the cluster mass, $M_{\rm pred}(a,b)$.  The results of this test are presented in the right column of \Crefrange{fig:star_DD_grad}{fig:ypar_DD_grad}, in the form of images of $\Delta M_{\rm pred}(a,b)$, the change in predicted mass when pixels in the neighbourhood of $(a,b)$ are masked.  Pixels that contribute significantly to the original mass estimate will produce a negative $\Delta M$ when masked.

For the stellar images, masking the central galaxy reduces the predicted mass dramatically, as we might expect. Beyond the central galaxy, the effects are much less clear.  There is some mild anti-correlation between the masked image and the DD image, as we might expect, but these are lower level effects compared to the central region.

For the X-ray-based tracers, the mask analysis shows that the outskirt of the X-ray data, where the signal gradient is largest, appears to be the most decisive feature the CNN triggers on.  For Compton $y$ images, the mask analsis shows that the central region plays important role but it fails to capture the details of the cluster. \ziangtxt{In summary, the mask method generally agrees with the DD analysis but is less informative, probably because the mask removes a fairly large region with details about cluster structure.}

As with the DD analysis, the CNN seems to be relatively insensitive to central regions.  \ziangtxt{We attribute this to the observation that the signal in the central region is more scattered with respect to cluster mass \citep{mantz2018centre, maughan2007lx}.}  We quantify this by calculating the correlation between the true mass and $F_{\rm cent}$ on the one hand, and $F_{\rm ring}$ on the other, where $F_{\rm cent}$ is the integrated X-ray signal in the range $r < 0.15R_{200}$ and $F_{\rm ring}$ is the integrated signal in the range $0.15R_{200} < r <R_{200}$.  The former has a correlation coefficient of 0.58 while the latter is 0.94 (for both Fxs and Fxb).

\section{Conclusion}
\label{sec:Conclusion}

We construct and train a set of convolutional neural networks (CNNs) to predict galaxy cluster masses, and test the network using a cluster catalogues derived from the BAHAMAS hydrodynamical simulations.  The clusters used in our study range in mass from $10^{12.7}$ to $10^{14.8}M_{\odot}$.  Using the simulation database, we generate mock data sets of stellar mass, soft X-ray flux, bolometric X-ray flux, and Compton $y$-parameter images as input, and train 4 single-channel networks on each of these observables independently. Each network has 3 convolutional layers and 3 pooling layers for feature extraction, followed by 5 fully-connected layers. We also construct a multi-channel network that takes all 4 data sets as simultaneous input. The multi-channel network is configured to run the 4 single-channel feature extraction sections independently. The output is then concatenated and processed by 6 fully-connected layers.  We train each network with 4800 randomly-selected cluster images and validate our training using $\sim$1600 validation images. The pre-trained network is then tested with  $\sim$1600 test images.

Our results are presented in \S\ref{sec:Results}.  All 5 of the networks successfully learn to predict cluster masses from mock data images.  In the mass range $10^{13.25} M_{\odot} < M < 10^{14.5} M_{\odot}$, our networks predict the true mass with a mean bias that is of order of 1\%. Outside of this range, our networks tend to over-predict the mass of low-mass clusters and under-predict the mass of high-mass clusters, \ziangtxt{which reflects a tendency towards mean.}  The per-cluster {\em rms} scatter is $\sim$15\%, with the Compton $y$ parameter and soft X-ray networks giving modestly lower scatter than the rest.  This performance is better than X-ray and tSZ-based analysis like \citet{zhang2008locuss} and \citet{bleem2015galaxy} while comparable to the weak lensing analysis of real data reported by \citet{umetsu2014clash}.  However, we note that weak lensing profiles bear richer mass information than the tracers we study, and current weak lensing studies focus on higher-mass clusters.  Future work applying CNNs to simulated weak lensing images would be needed to make a fairer comparison.

\citet{henson2016impact} estimate cluster masses in the BAHAMAS simulation by fitting the weak lensing profiles of all particles with empirical models.  They find a comparable mass bias to ours, however, they don't include noise, so the results are not directly comparable. So we conclude that our method is more accurate than previous methods which used the same hydrodynamical simulation, although the overall precision does not improve significantly. \ziangtxt{Although we have considered realistic systematics that could affect the performance of CNN, including beam smoothing, instrumental noise and additional structure along the line-of-sight. Particularly, by introducing fore- and background correlated signals, the mass scatter gets higher by $\sim 2\%$. We want to emphasize that it is important to include such systematics in future deep learning-related studies concerned with galaxy clusters mass estimates.} We also note that the data set in our analysis is still idealised compared to real observational data. 

We use two diagnostics to interpret the performance of our trained networks.  Both of them aim to identify image features that `trigger' the network to reach a particular conclusion.  The stellar mass CNN clearly detects galaxies and takes them into account when predicting cluster masses.  The gas-based CNNs apparently trigger on the shape and alignment of the gas, but the details are elusive. For example, the X-ray-based CNNs treat the cluster outskirts more importantly than the central region (in agreement with \citet{ntampaka2018deep}). The reason might be that cluster cores are known to be significantly scattered with mass, so the neural networks choose to ignore the central region for an optimal prediction. \ziangtxt{Similar future machine learning work could take this fact into account and down weight the central part by hand (by cutting out central region in preprocessing, for example).} 

This paper demonstrates a new approach to measuring galaxy cluster masses, a key parameter for understanding the origin and evolution of large scale  structure in the universe.  We show that a CNN is capable of recovering cluster masses directly from images of observable signals, despite the presence of substructure and noise. Our method does not require a physical model, however, it does require that one be able to simulate realistic clusters and systematic measurement errors reliably.  Future work might aim to train networks by combing data from different simulations, or even from real data.  We caution that neural networks are notoriously difficult to interpret, so future work should aim to better understand the behavior of hidden layers of the network.

\section*{Acknowledgements}

We thank Dr. Shirley Ho, Nayyer Raza, Riley Hill, Dr. Tilman Tr\"oster and Qiufan Lin for fruitful discussions. The calculations were performed on Compute Canada nodes (\url{www.computecanada.ca}).  AJM received funding from the European Union's Horizon 2020 research and innovation programme under the Marie Sk\l{}odowska-Curie grant agreement No. 702971. This work is financially supported by University of British Columbia and NSERC.  This project has received funding from the European Research Council (ERC) under the European Union's Horizon 2020 research and innovation programme (grant agreement No 769130).





\bibliographystyle{mnras}
\bibliography{main}


\bsp
\label{lastpage}
\end{document}